\documentclass[a4paper]{article}
\usepackage{graphicx}
\usepackage{amsmath}

\input{tcilatex}

\begin{document}

\title{Instabilities during the evaporation of a film:\\
non glassy polymer + volatile solvent}
\author{P.\ G.\ de Gennes \\
Coll\`{e}ge de France, 11 place Marcelin Berthelot\\
75231 Paris Cedex 05, France\\
e.mail: pgg@espci.fr}
\maketitle

\begin{abstract}
We consider solutions where the surface tension of the solvent $\gamma _{s}$
is smaller than the surface tension of the polymer $\gamma _{p}$. In an
evaporating film, a plume of solvent rich fluid, then induces a local
depression in surface tension, and the surface forces tend to strengthen the
plume. We give an estimate (at the level of scaling laws) for the minimum
thickness $e^{\ast }$ required to obtain this instability.\ We predict that
\ \ a) $e^{\ast }$ is a decreasing function of the solvent vapor pressure $%
p_{\text{v}}(e^{\ast }\sim p_{\text{v}}^{-1/2})$ \ \ b) $e^{\ast }$ should
be very small (\TEXTsymbol{<} 1 micron) provided that the initial solution
is rather dilute.\ \ \ c) the overall evaporation time for the film should
be much longer than the growth time of the instability. The instability
should lead to distortions of the free surface and may be optically
observable.\ It should dominate over\ the classical Benard-Marangoni
instability induced by cooling.

\medskip

\textit{PACS\ numbers}: 68.60Bs; 68.45Da; 6810 Jy.

\medskip

\textit{Shortened version of the title}:

\medskip

\ \ \ \ \ \ \ \ \ \ \ \ \ \ \ \ \ \ \ \ \ \ \ \ \ \ \ {\large Film
evaporation}
\end{abstract}

\section{Introduction}

Spin casting of polymer films from solutions is an important practical
process. The intrinsic state of these films after casting raises a number of
questions -especially if they are thin \cite{reiter}. In most cases, we are
dealing with materials such as polystyrene, or polymethyl metacrylate, which
are glassy in the final state: these glassy features complicate the process
enormously.

In the present pages, we consider a simpler case: \ \ \ \ a) the dry polymer
is assumed to remain fluid at room temperature (for instance, it could be a
silicone oil) \ \ \ \ b) the polymer is assumed not to adsorb on the free
surface of the solutions (this corresponds to $\gamma _{p}$\TEXTsymbol{>}$%
\gamma _{s}$).\ We shall see that certain convective instabilities should
occur during evaporation: one driven by concentration effects, and one
driven by thermal effects (Benard Marangoni). We find that the concentration
effects should usually be more important.

In section 2, we give a simple description of the classical evaporation
process in a mixed film, associated with a diffusion flow of the solvent.
Many more precise discussions of this process exist in the literature \cite
{saby dubreuil}, but the present crude picture is enough to estimate the
concentration drop $\psi _{d}-\psi _{u}$ between the bottom plate ($\psi
_{d} $) and the upper free surface ($\psi _{u}$).\ This drop is the
essential control parameter for the onset of a convective instability
(section 3).

This instability requires low viscosities: indeed the films\ are initially
dilute and of low viscosity.\ We discuss the dilute limit in section 4.

Ultimately, we discuss another possible source of instability related to
thermal effects: evaporation cools the outer surface. This leads to a
classical Benard Marangoni instability \cite{guyon}.\ We show in the
appendix that, (when $\gamma _{p}>\gamma _{s}$), concentration effects
should dominate over thermal effects.

All our discussion is restricted to the level of scaling laws. This is not
unreasonable: even without instabilities, the concentration profiles in the
evaporation problem are quite complex, and depend on many details: an exact
hydrodynamic calculation of the thresholds in this situation would be purely
numerical, and not very informative.

\section{Steady evaporation profiles}

The volume fraction of solvent $\psi (z)$ decreases when we move from the
bottom plate ($z=0$, $\psi =\psi _{d}$) to the free surface ($z=e(t),\psi
=\psi _{u}$).\ At the free surface, we assume instant equilibrium: the
partial gas pressure of the solvent is equal to the equilibrium gas pressure 
$p_{e}(\psi _{u})$, corresponding to $\psi _{u}$.\ In practice, we shall be
concerned mainly with dilute solutions and replace (very roughly) $%
p_{e}(\psi _{u})$ by the vapor pressure of pure solvent $p_{\text{v}}(T)$.
Above the free surface, the solvent molecules diffuse in air, with a certain
diffusion coefficient $D_{a}$.\ In practice, the air transport involves both
diffusion and convection, dependent on various noise sources in the
experimental room. As is often done, we assume that all this can be
described as diffusion through a boundary layer of fixed thickness $\ell $ ($%
\sim $ $1mm$).\ The number density of solvent molecules in the gas $\gamma
_{G}(z)$ thus drops linearly from the value associated to $p_{\text{v}}$:

\begin{equation}
\gamma _{G}(z=e)=\frac{p_{\text{v}}}{kT}  \label{eq1}
\end{equation}

to 0 at the upper end ($z=e+\ell $).

The diffusion current in the film (number of molecules per unit area and per
second) is:

\begin{equation}
W=D_{coop}\frac{\psi _{a}-\psi _{u}}{a^{3}e}  \label{eq2}
\end{equation}

where $D_{coop}$ is the cooperative diffusion coefficient of the solution 
\cite{pgg}, and we ignore the $\psi $ dependence of $D_{coop}$.\ The same
flux is found \ just above the free surface:

\begin{equation}
W=D_{a}\frac{\gamma _{G}(e)-\gamma _{G}(e+\ell )}{\ell }=D_{a}\frac{\gamma
_{G}(e)}{\ell }  \label{eq3}
\end{equation}

Our scaling estimate for the diffusion constant of solvent in air is $%
D_{air}\sim $v$_{th}\lambda $, where v$_{th}$ is a thermal velocity for a
solvent molecular, and $\lambda $ is a mean free path, inversely
proportional to the air pressure $p_{a}$.\ This gives ultimately:

\begin{equation}
D_{air}\cong \frac{\text{v}_{th}}{a^{2}}\,\frac{kT}{p_{a}}  \label{eq4}
\end{equation}

where $a$ is the size of a solvent molecule.

Combining eqs(\ref{eq1}-\ref{eq4}), we arrive at:

\begin{equation}
\psi _{d}-\psi _{u}\sim \frac{e}{\ell }\frac{a\text{v}_{th}}{D_{coop}}\frac{%
p_{\text{v}}}{p_{a}}  \label{eq5}
\end{equation}

Typically, taking \ $e=1\mu ,$ \ \ $\ell =1mm,$ \ \ $a$v$%
_{th}=10^{-3}cm^{2}/\sec ,$ \ \ $D_{coop}=10^{-6}cm^{2}/\sec ,$ and $\ \ p_{%
\text{v}}=0.1p_{a}$: we obtain $\psi _{d}-\psi _{u}\sim 0.1.$ Eq. (\ref{eq5}%
) is important, because it defines the driving force for the convective
instability to be discussed in the next section.

\section{Concentration}

The principle is shown on fig.\ref{fig1}. During the unperturbed evaporation
process, we saw that there is a significant difference between the solvent
fractions $\psi _{d}$ and $\psi _{u}$ (down and up) at the supporting
surface and at the free surface. We now assume that a roll instability is
superposed on the diffusion flux.\ We can visualise the system as a set of
rolls of size $e$, or as a set of plumes carrying extra solvent from down to
up.\ The result is a slight difference in solvent concentration between
point A ($\psi _{A}$) and point B ($\psi _{B}$).\ We estimate this
difference by a balance between convection and diffusion:

\begin{equation}
D_{coop}\frac{\psi _{A}-\psi _{B}}{e}\cong V(\psi _{d}-\psi _{u})
\label{eq6}
\end{equation}

where $V$ is the convective velocity.\ The difference $\psi _{A}-\psi _{B}$
induces a gradient of surface tension:

\begin{equation}
\nabla \gamma \cong \frac{\gamma B-\gamma A}{e}=\frac{-\gamma _{s}^{\prime }%
}{e}(\psi _{A}-\psi _{B})  \label{eq7}
\end{equation}

where $\gamma _{s}^{\prime }$=$d\gamma /d\psi $ is assumed negative: the
polymer has a higher surface tension than the solvent.

\FRAME{fhF}{3.3451in}{1.9527in}{0pt}{}{\Qlb{fig1}}{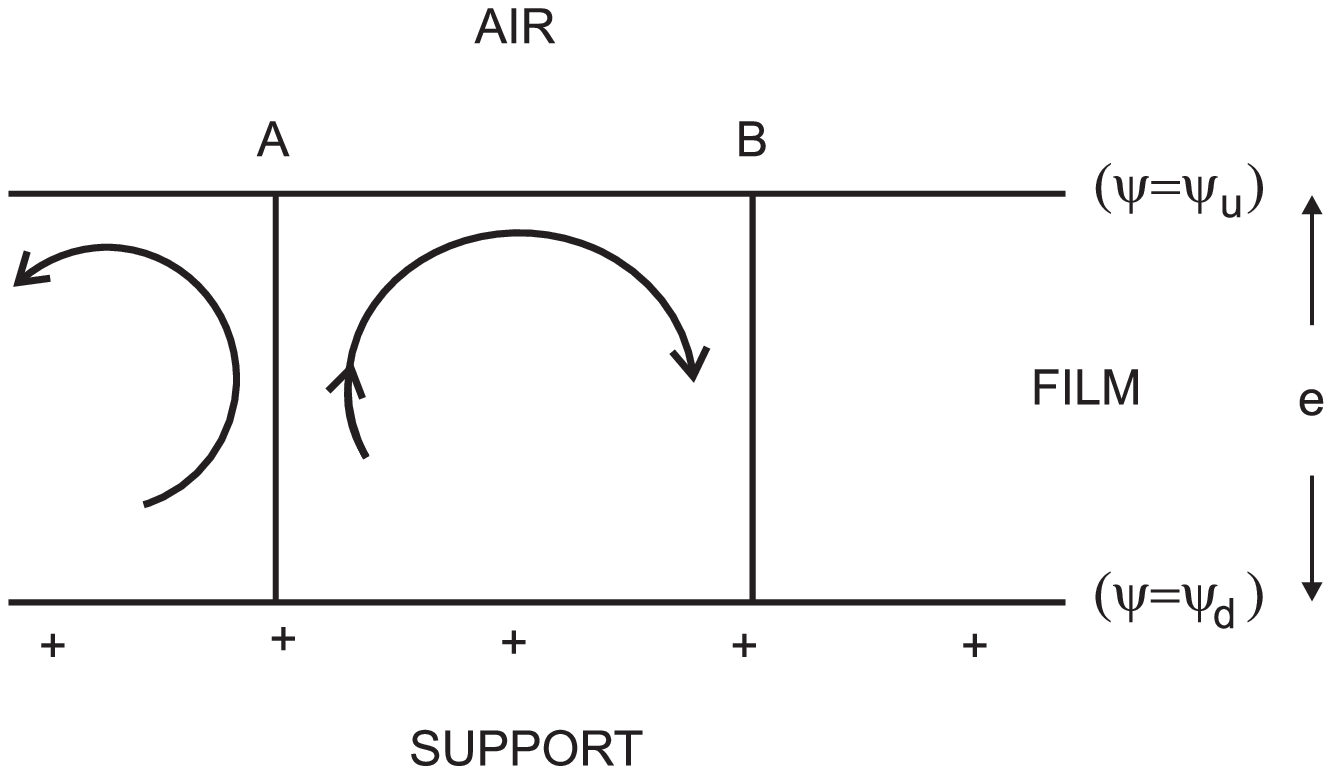}{\special%
{language "Scientific Word";type "GRAPHIC";maintain-aspect-ratio
TRUE;display "USEDEF";valid_file "F";width 3.3451in;height 1.9527in;depth
0pt;original-width 5.2805in;original-height 3.0649in;cropleft "0";croptop
"1";cropright "1";cropbottom "0";filename 'fig111.eps';file-properties
"NPEU";}}

Finally, we estimate $V$ by a balance between surface forces and viscous
stresses:

\begin{equation}
\nabla \gamma \sim \frac{\eta V}{e}  \label{eq8}
\end{equation}

where $\eta $ is the solution viscosity.

Combining eqs (\ref{eq6}, \ref{eq7}, $\ref{eq8}$) one arrives at a minimal
thickness:

\begin{equation}
e^{\ast }\cong \frac{\eta D_{coop}}{\left| \gamma _{s}^{\prime }\right|
(\psi _{d}-\psi _{u})}  \label{eq9}
\end{equation}

For rough estimates, we can write in the semi dilute regime:

\begin{equation}
D_{coop}=\frac{kT}{\eta _{s}\xi }  \label{eq10}
\end{equation}

where $kT$ is the thermal energy, $\eta _{s}$ the solvent viscosity, and $%
\xi $ the mesh size in the solution \cite{saby dubreuil}.

The result is:

\begin{equation}
e^{\ast }\cong \frac{kT}{\left| \gamma _{s}^{\prime }\right| \xi }\frac{\eta 
}{\eta _{s}(\psi _{d}-\psi _{u})}  \label{eq11}
\end{equation}

We can now combine eq. \ref{eq11} and eq. \ref{eq5}, arriving at:

\begin{equation}
(e^{\ast })^{2}\sim \frac{kT}{\left| \gamma ^{\prime }\right| }\frac{\ell
\eta }{\xi \eta _{s}}\frac{D_{coop}}{a\text{v}_{th}}\frac{p_{a}}{p_{\text{v}}%
}  \label{eq12}
\end{equation}

\section{Discussion}

\qquad a) Role of the solvent vapor pressure $p_{\text{v}}$: we see on eq. (%
\ref{eq12}) that high values of $p_{\text{v}}$ lead to low values of $%
e^{\ast }$: the instability (which occurs for $e>e^{\ast }$) is favored.

\qquad b) Role of the solution viscosity $\eta $: if we were dealing with
concentrated polymer films, $\eta /\eta _{s}$ would be very high, and $%
e^{\ast }$ would be prohibitively large.\ For instance, if we choose $\left|
\gamma ^{\prime }\right| =10mJ/m^{2},$ \ \ $\xi =10nm,$ \ \ $\eta /\eta
_{s}=10^{6},$ \ \ $\ell =1mm,$ \ \ and $D_{coop}/$v$_{th}a=10^{-3},$ we
arrive at $e^{\ast }\sim 0.2mm.$ However, we must not forget that our films
always start from a rather dilute state, with $\eta /\eta _{s}\sim 1.$ The
instability will grow \textit{during the early stages of evaporation}. If we
now switch to $\eta /\eta _{s}\sim 1$, we are led to $e^{\ast }=200$
nanometers. Thus (even if there are large numerical prefactors in eq.\ (\ref
{eq12})) we expect to find the instability in films around 1 micron.

\qquad c) To substantiate this prediction, we must show that the evaporation
time $\tau _{e\text{v}}$ is longer than the growth time ($\tau _{g}$) of the
instability.\ We estimate $\tau _{c\text{v}}$ as follows.

The rate of thinning is given by the flux $W$ of eq. (\ref{eq2}):

\begin{equation}
\frac{de}{dt}=-a^{3}W  \label{eq13}
\end{equation}

Using also eq. (\ref{eq2}), this gives an evaporation rate:

\begin{equation}
\frac{1}{\tau _{e\text{v}}}=-\frac{1}{e}\frac{de}{dt}=\frac{D_{coop}}{e^{2}}%
(\psi _{d}-\psi _{u})  \label{eq14}
\end{equation}

Lest us now consider the growth time of the instability $\tau _{g}$.\ For
simplicity, we shall focus our attention here on rather thin films, where
inertial effects can be neglected. We can then rewrite a time dependent
equation for the modulation $\psi _{m}\equiv \psi _{A}-\psi _{B}$ in the
form:

\begin{equation}
\frac{\partial \psi _{m}}{\partial t}=-D_{coop}q^{2}\psi _{m}+qV(\psi
_{d}-\psi _{u})  \label{eq15}
\end{equation}

where $q$ is the horizontal wave vector of the modulation ($q\sim e^{-1}$).\
We also have, as in eq. (\ref{eq8}):

\begin{equation}
\eta qV=q\left| \gamma ^{\prime }\right| \psi _{m}  \label{eq16}
\end{equation}

At threshold, the two terms in eq. (\ref{eq15}) balance each other.\ We now
consider situations definitively above threshold, where the convection term
in (15) is dominant.\ Using eq. \ref{eq16}, this leads to:

\begin{equation}
\frac{1}{\tau _{g}}\cong \frac{\left| \gamma ^{\prime }\right| }{\eta e}%
(\psi _{d}-\psi _{u})  \label{eq17}
\end{equation}

Comparing eqs (\ref{eq17}) and (\ref{eq14}) we find:

\begin{equation}
\frac{\tau _{e\text{v}}}{\tau _{g}}\cong \frac{\left| \gamma ^{\prime
}\right| e}{\eta D_{coop}}\cong \frac{\left| \gamma ^{\prime }\right| e\xi }{%
kT}  \label{eq18}
\end{equation}

Taking a dilute system, with $\left| \gamma ^{\prime }\right| =10mJ/m^{2},$
\ \ $e=1micron,$ \ \ $\xi =10nm,$ we find $\tau _{e\text{v}}/\tau _{g}\sim
2.10^{4}.\;$Thus, the instability should grow easily before complete
evaporation -during a time interval $\tau _{g}<<\tau _{e\text{v}}$, where
the viscosity is low and $e^{\ast }$ is small.

Another estimate\ for $\tau _{g}$, can be written down, when inertia is
dominant (and eq. \ref{eq16} is modified).\ But the conclusion remains the
same for most practical film thicknesses.

\qquad d) There are a number of details which are not included in our
discussion: for instance, the dependence of the surface tension $\gamma
(\psi )$ on solvent concentration in the dilute limit ($\psi \rightarrow 1$)
has not been considered. In fact, we know that when the polymer surface
tension is higher than the solvent surface tension, there is a depletion
layer of thickness $\xi $ near the free surface, and the surface tension has
the following scaling form \cite{saby dubreuil}:

\begin{equation}
\gamma (\psi )-\gamma (1)\cong \frac{kT}{\xi ^{2}}\cong \frac{kT}{a^{2}}%
(1-\psi )^{3/2}  \label{eq19}
\end{equation}

Thus, the parameter $\left| \gamma ^{\prime }\right| $ of eq. (\ref{eq7}) is
really $\psi $ dependent:

\begin{equation}
\left| \gamma ^{\prime }\right| \sim \frac{kT}{a^{2}}(1-\psi _{u})^{1/2}
\label{eq20}
\end{equation}

But, for reasonable values of $\psi _{u}$ (e.g. $\psi _{u}=0.9$) this brings
in only minor corrections on $e^{\ast }$ -proportional to (1-$\psi _{u}$)$%
^{1/4}.$

\section{Conclusions}

All our discussion ignored numerical coefficients, and we know how they can
be important (e.g. for the onset of thermal convection in the Rayleigh
problem). But some conclusions do emerge:

1) The convective instability due to concentration gradients in the film
should show up for non glassy polymer films (of thickness $e>e^{\ast }).$

2) It takes place early: when the film is still dilute.

3) It will induce a certain surface roughness.\ To compute the amplitude of
the surface undulations resulting from the rolls is a delicate task and will
not be attempted here.

\bigskip

\textit{Acknowledgments}: we benefited from discussions with G.\ Reiter, S.\
Kumar and F.\ Brochard-Wyart.

\section{Appendix}

We now ignore all concentrations effects, and focus on fig.\ \ref{fig2}.
Here, we have a thermal plume, terminating at point A, and we estimate the
temperature difference by a balance of heat fluxes:

\begin{equation}
D_{t}\frac{T_{A}-T_{B}}{e}=V(T_{d}-T_{u})  \tag{A1}
\end{equation}

\FRAME{fhF}{3.3537in}{1.695in}{0pt}{}{\Qlb{fig2}}{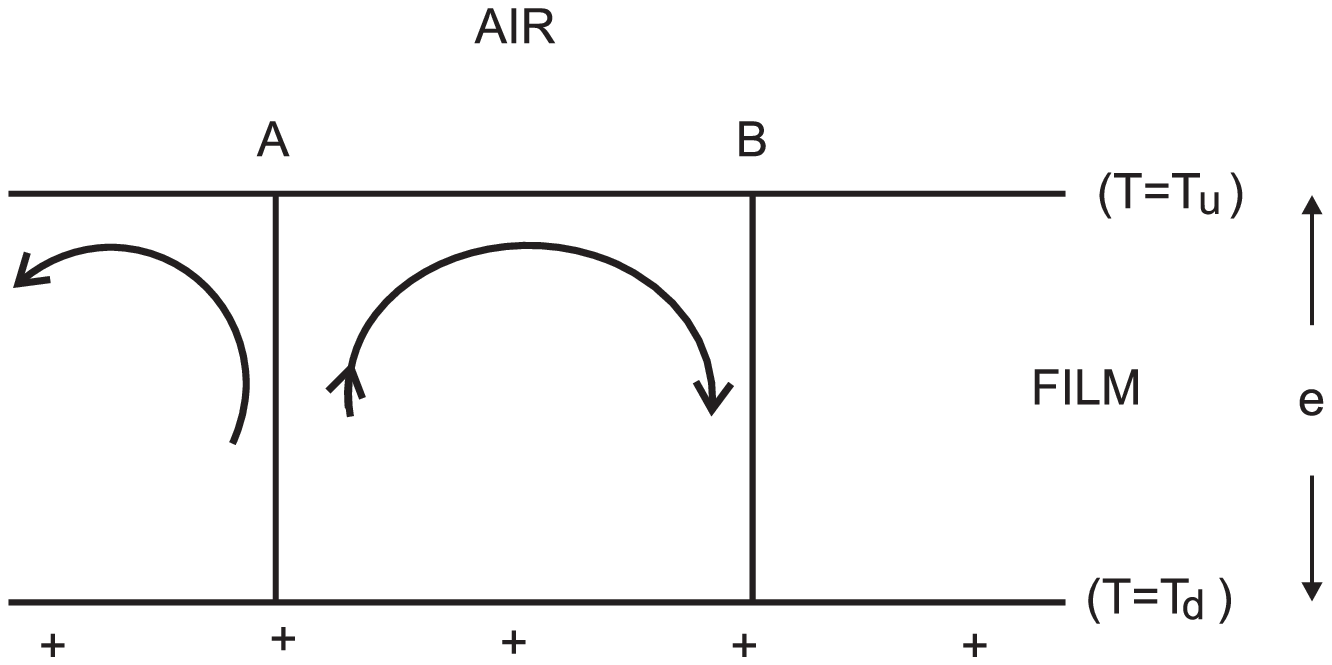}{\special%
{language "Scientific Word";type "GRAPHIC";maintain-aspect-ratio
TRUE;display "PICT";valid_file "F";width 3.3537in;height 1.695in;depth
0pt;original-width 5.2805in;original-height 2.6463in;cropleft "0";croptop
"1";cropright "1";cropbottom "0";filename 'fig23.eps';file-properties
"NPEU";}}

$D_{t}$ is the thermal diffusion coefficient of the liquid. The difference $%
T_{d}-T_{u}$ is due to evaporation and cooling at the upper surface. The
vertical solvent flux towards the air is:

\begin{equation}
J_{s}\cong D_{coop}\frac{\psi _{d}-\psi _{u}}{e}  \tag{A2}
\end{equation}

This corresponds to an outward energy flux:

\begin{equation}
J_{E}=U_{\text{v}}J_{s}  \tag{A3}
\end{equation}

where $U_{\text{v}}$ is the heat of evaporation (per unit volume of liquid).

Balancing $J_{E}$ against a vertical heat flux (from the support to the
outer surface), we obtain:

\begin{equation}
C_{p}D_{t}\frac{T_{d}-T_{u}}{e}\cong J_{E}  \tag{A4}
\end{equation}

Here, $C_{p}$ is the specific heat per unit volume (assumed constant), and $%
D_{t}$ a thermal diffusion coefficient in the liquid.

Finally, we have a stress balance similar to eq.\ \ref{eq8}, giving:

\begin{equation}
\eta V\cong -\gamma _{T}^{\prime }(T_{A}-T_{B})  \tag{A5}
\end{equation}

where: 
\begin{equation}
\gamma _{T}^{\prime }\equiv \frac{d\gamma }{dT}<0  \tag{A6}
\end{equation}

Combining eqs (A2-A5), we arrive at a threshold thickness:

\begin{equation}
e^{\ast \ast }=\frac{\eta }{\left| \gamma _{T}^{\prime }\right| }\frac{%
D_{t}^{2}}{D_{coop}}\frac{C_{p}}{U_{\text{v}}}\frac{1}{\psi _{d}-\psi _{u}} 
\tag{A7}
\end{equation}

We can now compare the two processes, however:

\begin{equation}
\frac{e^{\ast \ast }}{e^{\ast }}=\frac{\gamma _{s}^{\prime }}{\gamma
_{T}^{\prime }}\frac{C_{p}}{U_{\text{v}}}\left( \frac{D_{t}}{D_{coop}}%
\right) ^{2}  \tag{A8}
\end{equation}

The factors $\gamma _{s}^{\prime }C_{p}/\gamma _{T}^{\prime }U_{\text{v}}$
combine to give a constant of order unity, and the major features are
related to the diffusion coefficients. Typical values are $D_{coop}=10^{-6}$%
cm$^{2}/\sec ,$ $D_{t}=10^{-3}$cm$^{2}/\sec $, and we thus expect $e^{\ast
\ast }>>e^{\ast }$: for films with $\gamma _{p}>\gamma _{s},$\ thermal
processes should be dominated by concentration processes.

\end{document}